\def\comment#1{}

\newcommand{\ve}[1]{\mathbf{#1}}

\newcommand{\beg}{\begin{eqnarray}}
\newcommand{\eee}{\end{eqnarray}}

\documentclass[prl,letterpaper,twocolumn,aps,epsf]{revtex4}
\usepackage{dcolumn}
\usepackage{amsmath}
\usepackage{graphicx}%
\def\cm#1{}

\def\Kappa{K}
\def\sp{|{\Psi}_p|}
\def\se{|{\Psi}_e|}

\begin{document}
\title{
Violation of the London Law and Onsager-Feynman quantization in multicomponent superconductors
       }

\author{Egor Babaev${}^{1,2}$ and N. W. Ashcroft${}^{1}$}

\affiliation{
${}^1$Laboratory of Atomic and Solid State Physics, Cornell University,  Ithaca, NY 14853 USA
\\ ${}^2$ Department of Physics, The Royal Institute of Technology 10691 Stockholm, Sweden
}

\begin{abstract}
\end{abstract}

\maketitle
\newcommand{\la}{\label}
\newcommand{\aaa}{\frac{2 e}{\hbar c}}
\newcommand{\Pfaff}{{\rm\, Pfaff}}
\newcommand{\kA}{{\tilde A}}
\newcommand{\G}{{\cal G}}
\newcommand{\cP}{{\cal P}}
\newcommand{\M}{{\cal M}}
\newcommand{\E}{{\cal E}}
\newcommand{\btd}{{\bigtriangledown}}
\newcommand{\W}{{\cal W}}
\newcommand{\X}{{\cal X}}
\renewcommand{\O}{{\cal O}}
\renewcommand{\d}{{\rm\, d}}
\newcommand{\bfi}{{\bf i}}
\newcommand{\e}{{\rm\, e}}
\newcommand{\bfx}{{\bf \vec x}}
\newcommand{\bfn}{{ \vec{\bf  n}}}
\newcommand{\bfs}{{\vec{\bf s}}}
\newcommand{\bfE}{{\bf \vec E}}
\newcommand{\bfB}{{\bf \vec B}}
\newcommand{\bfv}{{\bf \vec v}}
\newcommand{\bfU}{{\bf \vec U}}
\newcommand{\bfp}{{\bf \vec p}}
\newcommand{\f}{\frac}
\newcommand{\bfA}{{\bf \vec A}}
\newcommand{\non}{\nonumber}
\newcommand{\be}{\begin{equation}}
\newcommand{\ee}{\end{equation}}
\newcommand{\ba}{\begin{eqnarray}}
\newcommand{\ea}{\end{eqnarry}}
\newcommand{\bastar}{\begin{eqnarray*}}
\newcommand{\eastar}{\end{eqnarray*}}
\newcommand{\half}{{1 \over 2}}

{\bf
Two quite fundamental principles governing the response to rotation of 
single-component superfluids and superconductors, the London law \cite{London}  relating the angular velocity to a subsequently established magnetic field, 
and  the Onsager-Feynman quantization of superfluid velocity
 \cite{Onsager,Feynman} are shown to be violated in a two-component superconductor. The manifestation of the two principles normally involves the fundamental constants alone, but this no longer holds as is demonstrated explicitly for the projected liquid metallic states of hydrogen and deuterium at high pressures. The rotational responses of liquid metallic hydrogen or deuterium identify them as a new class of
dissipationless states; they also directly point to a particular experimental route for verification of their existence.
}

Non-classical response to rotation is a hallmark of quantum ordered
 states such as superconductors and superfluids. 
The rotational responses of all presently known single-component ``super" states 
of matter (superconductors, superfluids and supersolids) 
are largely described by two fundamental principles
and   fall into two categories according to whether the systems are composed of neutral or charged particles.
For superfluid systems composed of electrically neutral particles (liquids, vapors, or even solids \cite{Andreev}) and for slow rotations, a fraction of the system, the superfluid fraction, remains irrotational.
However in response to  rotation 
exceeding a certain critical
rotation frequency, the superfluid fraction comes into
rotation by means of vortex formation. 
Onsager and Feynman \cite{Onsager,Feynman} pointed out that the
superfluid velocity ({\bf v}) in these vortices in single-component systems
is quantized and the  circulation quantum $\Kappa$
depends only on particle's mass $m$ and Planck's constant $\hbar$: 
$\Kappa = (1/2\pi) \oint {\bf v} \cdot d {\bf l}= \hbar/m$. 
Here we mention that for  superfluids with e.g. $p$- wave symmetry of the order parameter which are also invariant
under simultaneous phase and spin transformations this quantization  is
modified  \cite{volovik2}.
We also mention that a special situation
occur in a multicomponent superfluid with a
dissipationless drag (Andreev-Bashkin effect) 
where a superfluid velocity of one condensate  can carry
superfluid density of another  \cite{AB1,nstar}.
Below however we will consider the mixture of 
charged condensates only with the simplest symmetry of the order parameter
coupled by a gauge field.

For systems composed of charged particles and which are also superconducting 
(electronic Cooper pairs in metals, or  protonic Cooper pairs in neutron stars)
 vortices are not induced by rotation; however, the rotational response of these systems is no less interesting.
 London showed that a uniformly rotating single component
superconductor 
generates a persistent current in a thin layer near its
surface, and this in turn produces a detectable magnetic field, the London
field \cite{London}.  London related this field to the rotation frequency,
${\bf \Omega}$,
according to ${\bf B} = -(2mc/e){\bf \Omega}$, where $m$
is the electron's mass, $e$ denotes it electric charge 
and $c$ is the speed of light.
This law is experimentally confirmed (see e.g. \cite{London1} and references therein).
 Of
crucial significance is the fact that
the experimentally observed
London Law involves only the exact values of the fundamental constants, and
not on materials properties specific to the superconductor (such as an effective mass
for electrons). This law also holds for electronic superconductors with $d$- and $p$-wave pairing
symmetry.

We here consider the responses to rotation of the projected novel quantum states of metallic hydrogen and metallic deuterium, two-component
 systems
 exhibiting off-diagonal long-range order. 
These are now the subjects of 
renewed experimental pursuit especially because of the recent breakthrough in artificial
diamond technology. The expectation of achieving  
static pressures in diamond anvil cells
perhaps exceeding the expected metallization pressure of hydrogen at low temperatures 
has now been raised. 
Liquid metallic states 
of hydrogen were predicted earlier  to exhibit Cooper pairing 
both in protonic and electronic channels \cite{NWA2};
however it should be noted that an even simpler
situation may occur in liquid metallic deuterium because
 deuterons are bosons and can undergo condensation without the need for
a pairing instability. 
Another possible system where such 
states may be realized is a hydrogen-rich alloy
where under extreme but experimentally accessible 
pressures both electrons and protons may be mobile in a crystalline lattice 
\cite{hydrides}. Finally a rotational response
similar to that discussed below would be present in solid
 metallic hydrogen or deuterium if it exhibits a metallic  equivalent of
supersolidity.
For brevity below we shall always refer to ``liquid metallic hydrogen (LMH)"
but it is important to keep in mind that the range of
potential applications is much wider, including recent discussions
of possible presence of several charged barionic condensates
in neutron stars \cite{Jones}.
The main motivation of our study is to identify an effect which can  provide a possible experimental
probe for the renewed experimental search for superconducting liquid metallic hydrogen.

It has been observed that because in these systems the charged condensates are replicated twice 
(e.g. coexistent electronic and protonic, or deuteronic, condensates) composite neutral superfluid modes exist \cite{frac,frac2}.
These cannot be classified as superconductors in the usual sense;
we will see below that also
the superfluid mode is quite  different from 
 superfluid modes in one-component neutral systems.
Previous studies of this state have, however,  mostly focused
on the reaction of the system to an applied magnetic field \cite{frac,Nature,MSF};
here our intention
is to study the reaction of the system to rotation. 
The composite superfluid and superconducting modes
in this system are inextricably intertwined and as we find below
this has  unusual manifestations 
in rotational response, which extend our general understanding
of quantum ordered fluids.

The general route to describe a two component superconductor is the
London (or hydrodynamic) approach. The sysem in this approach
is described by the following free energy:
\begin{equation}
F = \sum_{\alpha=e,p}  \frac{1}{2m_\alpha}|\Psi_\alpha|^2(\nabla\theta_\alpha \pm
e\ve{A})^2
+ \frac{(\nabla\times{\bf A})^2}{2}.
\nonumber
\end{equation}
Here, 
$\Psi_\alpha$ and $m_\alpha, \ (\alpha=e,p)$ denote electronic and protonic condensate
wave functions and corresponding masses and $\bf A$ is the vector potential.
In what follows 
$e$ stands for an electric charge of a Cooper pair
and we set $\hbar=1, c=1$.
In this work we focus on the effects caused
by the coupling to the gauge field and thus we do not consider
possible drag effects \cite{AB1}.
Nor do we consider  different
pairing symmetries.

This model can be rewritten 
 as \cite{frac}
\begin{eqnarray}
&&F =  \frac{1}{2} \frac{\frac{|\Psi_e|^2}{m_e}\frac{|\Psi_p|^2}{m_p}}{ \frac{|\Psi_e|^2}{m_e}
+\frac{|\Psi_p|^2}{m_p}} 
\bigl(
\nabla\bigl( \theta_e+\theta_p\bigr)
\bigr)^2+ 
\frac{1}{2} \frac{1}{\frac{|\Psi_e|^2}{m_e}+\frac{|\Psi_p|^2}{m_p}}
\times \, \, \, \, \, \, \, \, \nonumber \\ 
&&
\Bigl( \frac{|\Psi_e|^2}{m_e} \nabla\theta_e
 - \frac{|\Psi_p|^2}{m_p}\nabla\theta_p
-e{\bf A} \Bigl[\frac{|\Psi_e|^2}{m_e}+\frac{|\Psi_p|^2}{m_p}\Bigr]
\Bigr)^2
+ \frac{{\bf B}^2}{2}.
\label{GLS}
\end{eqnarray}
The first term here  displays no
 coupling to the gauge field and therefore
represents a neutral or superfluid mode  which is associated 
with co-directed flows of electronic and protonic Cooper pairs (with no net
charge transfer) \cite{frac}.
 The second term accounts for the superconducting
(or charged) sector of the model describing
electrical currents. 
In what follows, we denote a vortex 
with phase windings $(\Delta\theta_e=2\pi n_e,\Delta\theta_p=2\pi n_p)$
as $(n_e,n_p)$.

Let us begin with inspection of the composite 
neutral mode's 
response to rotation.
The simplest topological excitation
in the superfluid sector of the model [i.e. a simplest
vortex which has a nontrivial winding in the phase sum $(\theta_e+\theta_p)$]
 is a vortex with the windings of only of one of the  phases: 
($\pm 1,0$) or ($0,\pm  1$).
We note that since the first term in (\ref{GLS}) is symmetric with respect to 
electronic and protonic condensates, both the 
($1,0$)
vortex and 
($0,1$) vortex
have identical configurations of the  neutral composite (i.e.
consisting of both electrons and protons) superflow. 
The difference between these two vortices
lies only in the contribution to the second term
in (\ref{GLS}) representing the charged (superconducting) sector
of the model.

We first focus on a 
(0,1) vortex.
For  this case, the solution for vector potential $\bf A$ at 
distances from the core much larger than penetration length
is given by \cite{frac} $
|{\bf A}| = \frac{1}{|e|r}{\frac{|\Psi_p|^2}{m_p}}\left[{ \frac{|\Psi_p|^2}{m_p}+\frac{|\Psi_e|^2}{m_e}}\right]^{-1}$,
where $r$ is the distance from the core center.
The superfluid velocities of electrons and protons in such a vortex
at a large distance from the core are
$ {\bf v}_p = (\nabla\theta_p+e{\bf A})/m_p; \  {\rm and} \
{\bf v}_e = -e{\bf A}/m_e$.
An equilibrium of a rotating system is achieved when the quantity
$E_r=E-{\bf  M}\cdot{\bf \Omega}$ is minimal (${\bf \Omega}$ is the rotation frequency and
${\bf M}$ and E are the angular momentum and energy).
Observe that if a system nucleates a vortex (1,0)
then not only protons but also electrons 
contribute to the angular momentum whose magnitude is given by:
$|{\bf  M}|=|{\bf M}_p+{\bf M}_e|=
\int( m_p\sp^2 v_p+ m_e\se^2 v_e) r dV  
$

The superfluid velocity circulations 
for protons and electrons in a vortex $(0,1)$
are given by:
$\oint {\bf v}_{(e,p)} \cdot d {\bf l}= 2\pi \Kappa_{(e,p)} =
 2\pi {\frac{|\Psi_{(p,e)}|^2}{m_{(p,e)}}}\left[{ \frac{|\Psi_p|^2}{m_p}+\frac{|\Psi_e|^2}{m_e}}\right]^{-1} \frac{1}{m_{(e,p)}}
$.
From this  we observe that {in the two-component superconductor the Onsager-Feynman
quantization rule is violated: the superfluid velocity quantization
is fractional} and  the electronic and protonic circulation 
quanta $K_{{e,p}}$ depend not only on 
mass but also on densities according to:
\beg
\Kappa_e = \frac{\frac{|\Psi_p|^2}{m_p}}{ \frac{|\Psi_p|^2}{m_p}+\frac{|\Psi_e|^2}{m_e}} \frac{1}{m_e}; \ \
\Kappa_p = \frac{\frac{|\Psi_e|^2}{m_e}}{ \frac{|\Psi_p|^2}{m_p}+\frac{|\Psi_p|^2}{m_e}} \frac{1}{m_p}.
\label{kappa}
\eee
The quantization conditions (\ref{kappa}) holds also
for the vortex $(1,0)$
It has been argued previously that quantization of
 magnetic flux in LMH is also
fractional \cite{frac}.  The fractionalization of superfluid velocity quantization which we find here
 has, however, a different
pattern. To compare the fractionalization of magnetic flux quantum $\Phi_0=2\pi/e$
and the fractionalization of superflow quantization we
 introduce an angle $\beta$
as a measure of the ratio of the average consensates densities, as follows:
$\sin^2\bigl(\frac{\beta}{2}\bigr)=\frac{|\Psi_e|^2}{m_e}\bigl[\frac{|\Psi_p|^2}{m_p}+\frac{|\Psi_e|^2}{m_e}\bigr]^{-1}
; \ \cos^2\bigl(\frac{\beta}{2}\bigr)=\frac{|\Psi_p|^2}{m_p}\bigl[\frac{|\Psi_p|^2}{m_p}+\frac{|\Psi_e|^2}{m_e}\bigr]^{-1}$.
Let $K_{(e,p)}^0=1/m_{(e,p)}$ be the standard superflow circulation quantum in a 
one component neutral superfluid composite of particles with the masses of electronic and protonic
Cooper pairs correspondingly. The quantization fractionalization pattern in this notation is then
summarized in the Table 1.

The energy per unit length ${\cal E}$ of vortices
$(1,0)$ and $(0,1)$
contains a logarithmically divergent part arising from the first term in ({\ref{GLS}}) \cite{frac,frac2}:
\beg
&&{\cal E}\approx \pi 
\left[ \sin^4\left(\f{\beta}{2}\right)  \f{|\Psi_p|^2}{m_p} + \cos^4\left(\f{\beta}{2}\right)  \f{|\Psi_e|^2}{m_e}\right]\log\f{R}{a}, \nonumber 
\label{energy}
\eee
where 
$a$ is a cut-off length which depends on the core structure 
and $R$ is the distance from the vortex center to the
system boundary.
 The formation
of vortices in response to rotation is controlled by the neutral mode [i.e.
by the first term in (\ref{GLS})]. As discussed
above, the vortices $(1,0)$ 
and $(0,1)$
have the same neutral superflow but different contributions to the
second term in (\ref{GLS}). The  energetically preferred excitations  forming in response to rotation are 
therefore the
$(0,1)$ vortices
which carry a smaller fraction of $\Phi_0$. 
We remark that composite vortices of the type $(\pm 1, \mp 1)$ do 
not contribute to superfluid sector of the model and are irrelevant in 
this rotational physics.
On the other hand it is straightforward to show that the vortices
 $(\pm 1,\pm 1)$ 
are  unstable.
\begin{center}
\begin{table}
\begin{center}
\caption{\label{table} Fractionalization of superflow circulation 
and magnetic flux quanta. 
}
\begin{ruledtabular}
\begin{tabular}{rccc}
   vortex:  &  $(1,0)$ &  $(0,1)$ \\
\hline
magnetic flux & $\sin^2(\beta/2)\Phi_0$ & $-\cos^2(\beta/2)\Phi_0$ \\
electronic superflow circulation & $\cos^2(\beta/2)\Kappa^0_e$ & $\cos^2(\beta/2)\Kappa^0_e$ \\
protonic superflow  circulation & $\sin^2(\beta/2)\Kappa^0_p$ & $\sin^2(\beta/2)\Kappa^0_p$ \\
\end{tabular}
\end{ruledtabular}
\label{table}
\end{center}
\end{table}
\end{center}

If a vortex 
$(0,1)$ is 
now placed into a cylindrical system with radius $R$ and unit height 
the system acquires an angular momentum:
$|{\bf M}|=\pi R^2
\frac{\se^2}{m_e}\frac{\sp^2}{m_p}\left[ \frac{|\Psi_p|^2}{m_p}+\frac{|\Psi_e|^2}{m_e}\right]^{-1}
\left( {m_e} + {m_p} \right).$
Vortices form when
$E_r={\cal E}-{\bf M\cdot\Omega} <0$. This determines
 the critical rotation
frequency as
\beg
\Omega_c\approx\f{1}{R^2(m_e+m_p)} \log\f{R}{a} 
\label{Omegac}
\eee
We can make a rough estimate of critical frequency:
$\Omega_c \approx({m_e}/{m_p})({e^2}/{a_0})(a_0/R)^2\log(R/a)$,
where $a_0$ is the Bohr radius, which for a $100 \mu$ sample
is of order of 10Hz.
Though we deal with a composite superfluid mode
and fractional circulation quantization, the critical frequency
is approximately the same as it would be in 
liquid of Cooper pairs of neutral particles with a mass of the proton. 
However  the underlying physics 
is indeed quite different.
One circumstance is that besides 
 fractional quantization of circulation,  only a small fraction of the condensates
participates in the superfluid mode (its stiffness is
$\frac{\se^2}{m_e}\frac{\sp^2}{m_p}\left[ \frac{|\Psi_p|^2}{m_p}+\frac{|\Psi_e|^2}{m_e}\right]^{-1}$).
Another difference can be seen
by considering a similar system but composed of
two types of  particles  with equal masses and  charge. This also 
features a superfluid mode but no vortices can be induced by rotation (this
also applies to electronic superconductors where multicomponent
order parameters arises from non $s$-wave pairing symmetry).

A quite deep difference in the rotational physics in two component charged 
systems is 
 manifested  especially in the novel ``aggregate states" of vortex matter they should allow.
As discussed above, in the simplest case the rotating system forms a lattice of vortices
$(0,1)$ (see Fig. 1A).
 In this respect the main
difference between this system and an ordinary superfluid is that the 
rotation-induced vortices are also
carrying magnetic flux $\Phi=\cos^2(\beta/2)\Phi_0$. The most interesting
situation arises when a rotating system is also subjected to a magnetic
field. In this case the  possible states of vortex matter are
numerous and we 
will consider here  some particularly interesting
possibilities of novel states of ``vortex matter".

If a weak magnetic field is applied in a direction 
opposite to 
the field of rotation-induced vortices 
the superconducting sector of (\ref{GLS}) would try to minimize its energy 
by introducing  $(1,-1)$ vortices.
These vortices have no neutral superflow (electronic
and protonic currents are counter-directed) but carry one 
magnetic flux quantum \cite{frac}. However a vortex $(1,-1)$
is not stable in a lattice of $(0,1)$
vortices because it experiences an attraction to such vortices
within the range of the penetration length scale \cite{frac2}. A vortex
$(1,-1)$ therefore should
annihilate with a $(0,1)$ vortex,
resulting in a $(1,0)$ vortex state.
At length scales larger than the penetration length a vortex
$(1,0)$ has a  Coulomb
repulsive interaction with a vortex $(0,1)$ 
similar to interaction between two
$(0,1)$ vortices \cite{frac2}
and therefore under normal conditions
 will occupy a space in a rotation-induced lattice of $(0,1)$ vortices;
it can therefore be viewed as a ground state ``electronic vortex-impurity" in a ``protonic
vortex lattice" (see Fig. 1B). The  concentration of these ``vortex-impurities" 
depends on the applied magnetic field and
there are indeed many interesting possibilities for their orderings and  phase transitions.

\begin{center}
\begin{figure}[htb]
\centerline{\scalebox{0.18}{{\includegraphics{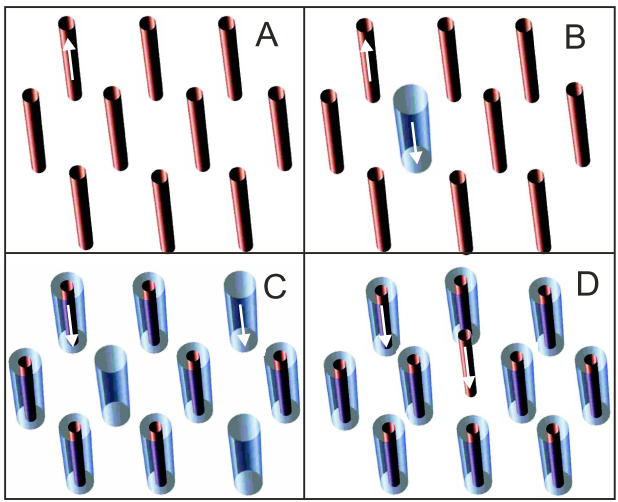}}}} 
\caption{
 {\sf A}:  A rotating 
two-component superconductor forms a lattice
 of
``protonic" vortices $(0,1)$ (red small tubes). The arrow 
denotes the direction of the self-induced magnetic field; {\sf B}:
When an  external field is applied in the direction opposite to the field
carried by rotation-induced vortices, the ground state 
posseses ``electonic vortices"
$(1,0)$ (blue tubes)
placed as ``impurities" in protonic vortex lattice; {\sf C, D}: 
In a strong magnetic field, and for
rotations just above the critical frequency, 
the ground state features either ``electronic vortex impurities"
in the lattice of composite vortices or ``interstitial" protonic vortices, 
depending on the applied field direction.}
\end{figure}
\end{center}

Consider next a situation with a stronger magnetic field
and with a 
rotation frequency just above $\Omega_c$. Then the dominant structure is a field-induced
lattice of composite vortices $(1,-1)$ (as
in the case of no rotation \cite{frac,Nature,MSF}). 
Here the energetically most favorable way to introduce a superfluid momentum-carrying
vortex is the substitution of one of the $(1,-1)$ 
vortices by a $(1,0)$ vortex (see Fig. 1C).
This vortex interacts repulsively with its neighbors, carries almost one
magnetic flux quantum, but also posseses angular momentum in the superfluid sector.
This vortex is therefore an ``elementary vortex-impurity" in a lattice of composite vortices.
Such a system should exhibit a number of novel phase transitions and vortex matter
states.
One such transition will occur
because there is a finite potential barrier for a ``vortex impurity" to jump
from one lattice site to another. At certain temperatures  the vortex impurities
should be able to move from one site to another freely. There is an analogy 
between ``light" vortices, which in the ground
state are co-centered with ``heavy" vortices 
and a system of Bose particles living on sites of a triangular lattice 
which is known to possess a supersolid state
(see for example \cite{BH1}). In this analogy the $z$-direction
in the vortex lattices plays the role of ``time" in the Bose system and the``heavy" vortices 
play the role of a robust ``lattice potential". There arises therefore 
the possibility of an interesting ``aggregate
state of vortex matter", analogous
to the supersolid state of Bose particles which 
in our case may be called a ``vortex supersolid".
This state should feature coexistence of crystalline order of vortices
and ``fluidity" of ``vortex impurities". Because of the fluid state of light vortices
the phase of the corresponding condensate will
be disordered in the $z$-direction which in turn should destroy the order in gauge-invariant
phase sum. For this reason the transition into this ``vortex supersolid"
phase has an important physical consequence, namely,
the disappearance of superfluidity of particles along
the rotation axis,
which constitututes the phase
transition from superconducting superfluidity to superconductivity
selectively along this direction.

In the case where the rotation direction is inverted, while the magnetic field unchanged,
the topological defects in the superfluid sector
which minimize  energy in a rotating frame are 
$(-1,0)$ and $(0,-1)$.
The former vortex subtracts almost one magnetic flux quantum and should be 
compensated by two $(1,-1)$ vortices.
In this  scenario a lattice should becomes
more dense (with a certain energy penalty). 
On the other hand there is a second possibility
to acquire angular momentum: the introduction of a $(0,-1)$ vortex.
This results in a different type of the energy penalty:
a vortex $(0,-1)$ interacts
via a screened potential with a composite vortex $(1,-1)$
but the interaction strength is much weaker than between
two composite vortices \cite{frac2}. Therefore such a vortex 
 can be introduced as an ``interstitial vortex defect" (see Fig 1D) which, for a range
of parameters, should be a
more energetically preferred way to acquire angular momentum 
than the first possibility. The ``light" vortices may form an ``interstitial vortex liquid" state, while
the co-centricity of light vortices with the lattice of heavy vortices
is controlled by a different energy scale. This is again 
a state with coexistent vortex crystalline order and
vortex defect fluidity and yet another example of a  ``vortex supersolid" 
which resembles the supersolid state of interstitial particles in crystals
discussed in \cite{Andreev}.

Finally, let us consider  the reaction of the superconducting
sector of the system to rotation.
It is important  to note that electronic and protonic Cooper pair momenta  depend on the same vector potential,
${\bf P_\alpha} \equiv \nabla\theta_\alpha= m_\alpha {\bf v}_\alpha + e_\alpha {\bf A}$ and hence 
${\bf A}=\f{{\bf P_\alpha} }{e_\alpha}-\f{m_\alpha}{e_\alpha}{\bf v}_\alpha$
(where $e_{(e,p)}=\pm e$).
Consider now the situation without an applied external field and low rotation frequencies,
so that there are no vortices (i.e.  $\Omega<\Omega_c$). Then taking the curl of the previous expression we arrive at the
constraint dictated by gauge-invariance:
\beg
\f{m_p}{e_p}\nabla\times{\bf v}_p=\f{m_e}{e_e}\nabla\times{\bf v}_e.
\label{const}
\eee
Let us consider first the  zero temperature case
when there is no normal component.
If  the condensate charges
entering the problem are opposite (as is indeed the case
for LMH) this equation
has a trivial solution: ${\bf v}_p={\bf v}_e=0$ i.e. at $T=0$ 
for  $\Omega<\Omega_c$ the condensates remain irrotational. However 
in the presence of a normal component with a net electric charge its rotation
produces an electric current so the superconducting component necessarily has to respond
(i.e. ${\bf v}_p={\bf v}_e=0$ can no longer be a stationary solution).
From (\ref{const}) it also follows that in contrast
to London's picture for ordinary superconductors \cite{London},
superconducting electrons and protons will not follow the rotation of normal component
because it would violate constraint  (\ref{const}). This dictates a counter-intuitive situation, namely that in response to slow rotation
the superconducting electrons and protons { can only  move in opposite 
directions and at different speeds}. Their superconducting velocities can be expressed
in the following form: ${\bf v}_\alpha=\gamma_\alpha {\bf  \Omega}\times {\bf r}$.
To find  $\gamma_\alpha$ we first observe that from the stationarity
requirement  we can obtain an
extra condition by equating the rotation-induced electric current of the normal component (multiplied by -1)
to the rotation-induced current response of superconducting sector subject to constraint (\ref{const}):
${\bf J}_s=(e_p\gamma_p|\Psi_p|^2+e_e\gamma_e|\Psi_e|^2) {\bf  \Omega}\times {\bf r}$.
From the overall electrical neutrality of the system it follows that the rotation-induced
normal current is ${\bf J}_n=-(e_p|\Psi_p|^2+e_e|\Psi_e|^2) {\bf  \Omega}\times {\bf r}$. Hence we find
\beg
{\bf v}_p=\f{|\Psi_p|^2-|\Psi_e|^2}{|\Psi_p|^2+\f{m_p}{m_e}|\Psi_e|^2 }{\bf  \Omega}\times {\bf r};
 \  {\bf v}_e=\f{|\Psi_e|^2-|\Psi_p|^2}{|\Psi_e|^2+\f{m_e}{m_p}|\Psi_p|^2 }{\bf  \Omega}\times {\bf r}.
\nonumber
\eee
To sustain these counter currents a rotating two-gap superconductor should generate in its bulk a vector potential
and hence  rotation induces a magnetic field:
\beg
{\bf B}_{rot}= \f{2}{e}\f{|\Psi_p|^2-|\Psi_e|^2}{\f{|\Psi_p|^2}{m_p}+\f{|\Psi_e|^2}{m_e} }{\bf  \Omega}.
\label{brot}
\eee
While in the bulk the superfluid electrons and protons have the velocities ${\bf v}_{e,p}$,
the field ${\bf B}_{rot}$ is generated by velocity variations in the layer near the system's edge with
the thickness of the penetration length  $\lambda=(e^2[{|\Psi_p|^2}/{m_p}+{|\Psi_e|^2}/{m_e} ])^{-1/2}$.
This follows from the equation for magnetic field variation, namely:
$-\lambda^2 \nabla^2 {\bf B}({\bf r}) + {\bf B}({\bf r}) ={\bf B}_{rot}$. Eq. (\ref{brot}) demonstrates a 
remarkable circumstance: the
London Law in the two-component superconductor is actually violated. The rotation-induced
field is {\it not a universal  function  of the fundamental constants  irrespective
of microscopic details}. Indeed it aquires a  dependence on densities.
 At temperatures just below superconducting transition for protons
 a rotating sample of radius $ R$
generates a magnetic flux of order of $R^2\Omega$ flux quanta  ($R$ given in cm and $\Omega$ in s$^{-1}$)
which could be detectable with modern SQUIDs even for samples as small as $10\mu$ 
rotating at $1Hz$ (we note that it is easier to achieve high pressures
in small samples which makes it a very convenient experimental probe).
 Going to a larger sample or higher rotation frequency
would even allow 
 measurement of rathio of the condensates densities and their
temperature dependences, as follows from (\ref{brot}).
And, of course, its absence
would even  rule out protonic superconductivity or deuteronic condensation.
It follows that a direct experimental route exists for the verification of 
this possible new class of dissipationless states.
\begin{center}
\begin{figure}[htb]
\centerline{\scalebox{0.3}{{\includegraphics{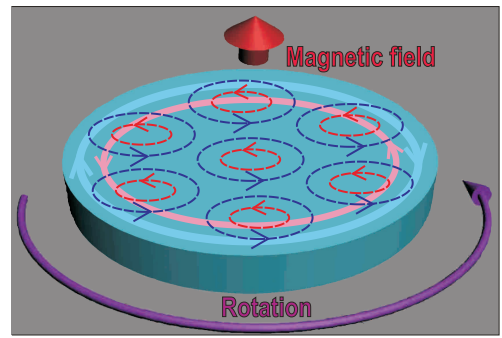}}}} 
\caption{ 
If a two-component superconductor is
 brought into  rotation two types of particle flow are generated:  for slow
 rotation  the two components (here electrons and protons) start circulating in opposite directions (schematically
denoted by thick red and blue arrows) inducing
a magnetic field $\bf B$ along the rotation axis. At 
 a faster rotation a second type of particle flow also appears 
in the form of quantum vortices.
Here electrons and protons
flow in the same direction (thin dashed lines). As discussed in the text in this picture the basic laws
governing rotational response of one-component quantum fluids: the Onsager-Feynman
superflow circulation and the London law are violated. 
 }
\end{figure}
\end{center}

Though this has been cast in terms of a possible failure of London's law
(otherwise rigorously applicable up to relativistic corrections in
electronic superconductors) the major issue discussed here might well be
viewed as a possible extension of the classifications of the rotational
responses of quantum fluids. Rotational response is a quintessentially
state-defining property of quantum fluids, and the one we find in LMH (as
summarized in Figure 2) is seen to be quite complex, and it involves both
co- and
counter-directed electrical currents, and in particular a current in the
direction opposite to rotation. This suggests a classification of the
projected liquid state of metallic hydrogen as a new quantum fluid, and
one which may be presenting considerable opportunity for new and emerging
physics.
From an
experimental point of view it now appears that the rotational response of
muticomponent superconductors may well offer the most direct probe, both
qualitative and quantitative, of the corresponding quantum orderings of 
hydrogen and deuterium in experiments in diamond anvil cells.
It may also be extendable to the ternary systems formed by addition
of further multivalent ions to the proposed ground state metallic fluids where under high pressure
protonic diffusive states may yet persist in the presence of periodic ordering
of the ions.

{\small
This work was supported by the National Science Foundation under Grants 
DMR-0302347 and DMR-0601461. 

Correspondence and requests for materials should be addressed to E.B.}

\end{document}